\documentclass{aa}
\usepackage{graphicx}
\usepackage[T1]{fontenc}

\usepackage{amsmath}
\usepackage{amssymb}

\usepackage{natbib}
\bibpunct{(}{)}{;}{a}{}{,}

\usepackage{caption}

\usepackage[figuresright]{rotating}

\begin{document}

\title{Weak lensing density profiles and mass reconstructions\\
  of the galaxy clusters Abell~1351 and Abell~1995
  \thanks{Based on observations obtained with CFH12K at the
    Canada-France-Hawaii Telescope (CFHT) which is operated by the
    National Research Council of Canada, the Institut des Sciences de
    l'Univers of the Centre National de la Recherche Scientifique and
    the University of Hawaii.}
}
\author{K. Holhjem\inst{1,2}
  \and M. Schirmer\inst{1,3}
  \and H. Dahle\inst{2}}

\offprints{K. Holhjem}

\institute{Argelander-Institut f\"ur Astronomie (AIfA), Universit\"at Bonn, 
  Auf dem H\"ugel 71, D-53121 Bonn, Germany\\
  \email{kholhjem@astro.uni-bonn.de}
  \and Institute of Theoretical Astrophysics, University of Oslo, P.O. Box
  1029 Blindern, N-0315 Oslo, Norway
  \and Isaac Newton Group of Telescopes, Calle Alvarez Abreu 70,
  E-38700 Santa Cruz de La Palma, Spain
}

\date{Received / Accepted }

\abstract{}
{The aim of the present work is to study the overall mass distribution
  of the galaxy clusters Abell~1351 and Abell~1995 using weak
  gravitational lensing. These clusters have got a very different mass
  structure and dynamical state, and are the two extremes from a
  larger sample of 38 X-ray luminous clusters of similar size and
  redshift.}
{We measure shear values of faint background galaxies and correct for
  PSF anisotropies using the KSB+ method. Two-dimensional mass maps of
  the clusters are created using a finite-field mass reconstruction
  algorithm, and 
  verified with aperture mass statistics. The masses inferred from the
  reconstructions are compared to those obtained from fitting
  spherically symmetric SIS- and NFW-models to the tangential shear
  profiles.
  We discuss the NFW concentration parameters in detail.}
{From the mass reconstructions we infer $M_{\rm 2D}(<r_{200})$-masses of
  $11.7\pm3.1\times10^{14}h_{70}^{-1}M_\odot$ and
  $10.5\pm2.7\times10^{14}h_{70}^{-1}M_\odot$ for Abell~1351 and Abell~1995,
  respectively. About $3\arcmin$ north-east of the main
  mass peak of Abell~1351 we detect a significant secondary peak in
  the mass reconstruction as well as by aperture mass statistics. This
  feature is also traced by clusters members selected by means of
  their \mbox{$V-I$} colour, and hence is likely a real sub-structure of
  Abell~1351. From our fits to the tangential shear we infer
  masses on the order of $M_{200}\sim8-9\times10^{14}h_{70}^{-1}M_\odot$
  (Abell~1351) and $M_{200}\sim5-6\times10^{14}h_{70}^{-1}M_\odot$
  (Abell~1995). The concentration parameters remain poorly
  constrained by our weak lensing analysis.
}
{}
\keywords{Gravitational lensing -- Cosmology: dark matter -- Galaxies:
  clusters: individual: Abell~1351 and Abell~1995} 

\titlerunning{Weak lensing analysis of Abell~1351 and Abell~1995} 
\maketitle

\section{Introduction}

Comprising the most massive gravitationally bound structures in
the Universe, galaxy clusters are essential in providing a deeper
understanding of the properties of dark matter. The recent papers
  covering $1$E~$0657-558$ (the Bullet Cluster) by \citet{cbg06} and
Abell~520 by \citet{mhb07}
demonstrate the importance of
gravitational lensing
for understanding the matter content of our Universe.
Lensing studies of galaxy clusters provide
a powerful way to identify large density peaks in the Universe,
independent of their baryonic content
\citep{msm07,mhe07,gs07,seh07,wdh06,csa06,dpl03,sch96}.

Neither the nature nor the dynamical state of the gravitating matter
affect the mass estimates obtained through gravitational
lensing. These mass measurements are only changed by gravitation and
the geometrical configuration between observer, lens, and
source. Although this makes lensing a unique tool,
such measurements can be biased by e.g. different mass
concentrations along the line-of-sight.

In this paper we analyse two clusters of galaxies
at intermediate redshift \mbox{($z=0.32$)}, Abell~1351 and
Abell~1995. These clusters are selected from a lensing study of
38 highly X-ray luminous galaxy clusters
\citep{dki02}. They were chosen for further
investigation as they represent the two extremes in this cluster
sample, regarding mass distribution and dynamical state. Re-observing
the clusters with the wide-field camera CFH12K at the
Canada-France-Hawaii Telescope (CFHT) provided a larger field of view
\mbox{($42\farcm2\times28\farcm1$)} than that employed by
\citet{dki02} \mbox{($18\farcm8\times18\farcm8$)}. This allows us to
map the clusters to larger radii than previously possible.

The KSB+ method \citep{ksb95,lk97,hfk98} is used to
recover the shear values of faint background galaxies in the
images. 
Using a finite-field reconstruction technique \citep[][SS01]{ss01} we 
derive two-dimensional mass maps, visualising the surface mass
distributions of Abell~1351 and Abell~1995. We also apply aperture 
mass statistics \citep[$M_{\rm ap}$;][]{sch96,seh07} to our data,
comparing the results to confirm mass peak detections.
Finally, by fitting predicted shear values from theoretical
models to the shapes of the lensed galaxies we estimate the cluster
masses.
We assume the
clusters to be spherically symmetric and to have density profiles
following either a singular isothermal sphere (SIS) or a
\citeauthor*{nfw97} \citetext{\citeyear{nfw97}, \citeyear{nfw95}; NFW}
model.

Data processing and analysis are carried out using mainly the {\tt
  IMCAT} software package\footnote{{\tt IMCAT} is developed by Nick
  Kaiser (kaiser@hawaii.edu),
{\tt http://www.ifa.hawaii.edu/$\sim$kaiser/imcat/}}, Kaiser's July
2005 version for Macintosh. {\tt IMCAT} is a tool specially designed
for weak lensing purposes, and is optimised for shape measurements of
faint galaxies. It processes both {\tt FITS} files and object
catalogues.

The outline of this paper is as follows. In Sect.~\ref{observations}
we summarise the observations and software used for our
study, as well as the data reduction. In Sect.~\ref{shear} we describe
the shear reconstruction, focusing on shape estimates and Point Spread
Function (PSF) corrections. In Sect.~\ref{mass} we present the
clusters' surface mass density maps and attempt to verify the detected
mass peaks using \mbox{$V-I$} colours and $M_{\rm ap}$.
By comparing the measured shear profiles to
theoretical expectations,
we model the lensing data in
Sect.~\ref{model}. Finally we present and discuss our results in
Sect.~\ref{discussion}, and our conclusions in
Sect.~\ref{conclusion}.

Throughout this paper we assume a \mbox{$\Lambda$CDM} cosmology,
with \mbox{$\Omega_M=0.3$}, \mbox{$\Omega_\Lambda=0.7$}, and
\mbox{$h_{70}=H_0/(70$ km s$^{-1}$ Mpc$^{-1}$)}. Errors are given on
the \mbox{$1\sigma$} level.

\section{Observations and data reduction}\label{observations}

\subsection{Data acquisition}

The galaxy clusters Abell~1351 and Abell~1995 are centred at the
positions
\mbox{$11^{\rm h}42^{\rm m}30\fs7\,+58\degr32\arcmin21\arcsec$} and
\mbox{$14^{\rm h}52^{\rm m}50\fs4\,+58\degr02\arcmin48\arcsec$},
respectively. They were observed with the $3.6\;{\rm m}$
Canada-France-Hawaii Telescope (CFHT) on the 4 nights of 7-11 May,
2000, using the wide-field CCD mosaic camera CFH12K. A total exposure
time of 5400s were obtained for both clusters, all in the
\emph{I}-band filter. However, due to seeing \mbox{$>1\arcsec$} three
exposures were rejected from the Abell~1351 data set, resulting in
total exposure times of $3600\;{\rm s}$ and $5400\;{\rm s}$ for
Abell~1351 and Abell~1995, respectively. This corresponds to a
$5\sigma$ limiting magnitude of $I=25.2\pm0.1$ for point sources
in both pointings. The seeing in the final co-added images
respectively equals $0\farcs95$ and $1\farcs15$ for Abell~1351 and
Abell~1995. The number density of the lensed background galaxies is
$16$ arcmin$^{-2}$ for both clusters, and the ellipticity
dispersion (after PSF correction) for each component is
$\sigma_g=0.43$ and $\sigma_g=0.51$ for Abell~1351 and Abell~1995,
respectively. The larger dispersion for Abell~1995 is explained by
the $20\%$ larger image seeing, which enlarges the PSF correction
factors and their uncertainties.

The CFH12K mosaic camera covers a field of
\mbox{$12{\rm k}\times8{\rm k}$} pixels in total, representing
an area of
\mbox{$42\farcm2\times28\farcm1$} on the sky. The pixel scale is
\mbox{$0\farcs206$} when mounted at the CFHT prime focus.

In addition we make use of the \emph{V}-band data obtained by
\citet{dki02} in order to verify neighbouring peaks
present in our two-dimensional mass maps described in
Sect.~\ref{mass}. These data were obtained at the $2.24\;{\rm m}$
University of Hawaii Telescope using the UH8K mosaic camera, covering
an area of \mbox{$4{\rm k}\times4{\rm k}$} pixels (re-binned
\mbox{$2\times2$}) mapping \mbox{$18\farcm8\times18\farcm8$} of the
sky. Each image has a total exposure time of $12\,600\;{\rm s}$,
resulting in an image depth comparable to our \emph{I}-band data, with
$5\sigma$ limiting magnitudes of $V=25.9\pm0.1$ and $V=25.8\pm0.1$ for
Abell~1351 and Abell~1995, respectively. Further details on the
reduction process and co-addition of the \emph{V}-band data can be
found in \citet{dki02}.

As Abell~1351 and Abell~1995 are both located at redshift $z=0.32$,
they have a similar correspondence between physical and angular scale,
given as \mbox{$1{\rm Mpc}=215\arcsec$}.

\subsection{Image processing}\label{imageprocessing}

\subsubsection{Pre-processing}\label{background}

To remove the bias level in each frame we used the mean value of
  the overscan region from the corresponding chip. The
flat-fielding was carried out using a master night time flat, made
from averaging 56 night time exposures; most of them so-called
``blank'' fields and all well displaced from each other. The fringing
that occurs in \emph{I}-band exposures is also captured in this type
of flat, and upon dividing the object exposures by it the fringes were
cleanly removed.
To estimate the background level in the exposures, we used the heights
of the minima of the sky level present to create a model for each
individual frame. After subtraction the median sky level was set to
zero.

As fringing is an additive effect and not a multiplicative one,
ideally the fringes should be subtracted. Since we had no twilight
flats available, standard defringing could not be performed. The
photometric error introduced
by division is negligible, as the amplitude of the fringes compared to the
sky background after flat-fielding was on the order of $2\%$.
However, since fringing acts mostly on small angular scales, its treatment 
will affect the shapes of the small and faint background galaxies used for 
weak lensing. To investigate this we obtained a set of 10 archival
images of the Deep3 field \citep{hed06}, taken with the Wide Field
Imager at the 2.2m MPG/ESO-telescope through their $I$-band
filter.
As the Deep3 field does not contain any massive clusters it is
very well suited for this test.
Two different coadded images were created. In the first case
the data were flat-fielded using twilight flats, and then a fringing
model was created from the flat-fielded data and subtracted. The
second coadded image was processed in the same way as our CFHT data,
i.e. the data were flat-fielded and fringe-corrected by division of a
night-time flat. We then measured the shapes of a common set of
$\sim12000$ galaxies in both images (see Sect.~\ref{shear}) and
created two mass reconstructions, using the same technique
and smoothing scale as for Abell~1351 and Abell~1995 (see
Sect.~\ref{mass}).
We find that the rms of the difference of the two mass maps is a
factor of 2.5 smaller than the noise of the individual mass maps,
mainly caused by the intrinsic ellipticities of background galaxies.
The effect in our CFH12K data is much smaller, as the CFH12K
\emph{I}-band filter has a blue cut-on at around 730 nm and a red
cut-off at 950nm. The ESO \emph{I}-band filter on the other hand opens
at 800nm and has no
cut-off on the red side. Hence the fringing amplitude in the
comparison data set from ESO is up to 5 times higher than in
CFHT. We conclude that our analysis of Abell~1351 and Abell~1995 is
not affected by our fringe correction.

\subsubsection{Masking}\label{masking}

The CFH12K mosaic contains some bad pixels and columns, in particular
two of the CCDs suffer from this. By using Nick Kaiser's ready made
CFH12K masks\footnote{\tt
  http://www.ifa.hawaii.edu/$\sim$kaiser/cfh12k/masks/} as global
masks, all bad areas were ensured to be ignored.
An additional patch of 219 bad columns in CCD00 was also added to the
global masks.
We did not make
further individual masks for each exposure, as most spurious
detections are filtered out during the astrometric calibration. In
addition, suspicious objects are rejected from the final object
catalogue by visual examination.

\subsubsection{Astrometric calibration}\label{astrometry}

Wide-field data typically do not have a simple relation between the
sky coordinates and those of the detector. A mapping from pixel
coordinates onto a planar projection of the sky needs therefore be
performed. We solved for this through a series of steps.

\begin{figure*}
  \includegraphics[width=17cm]{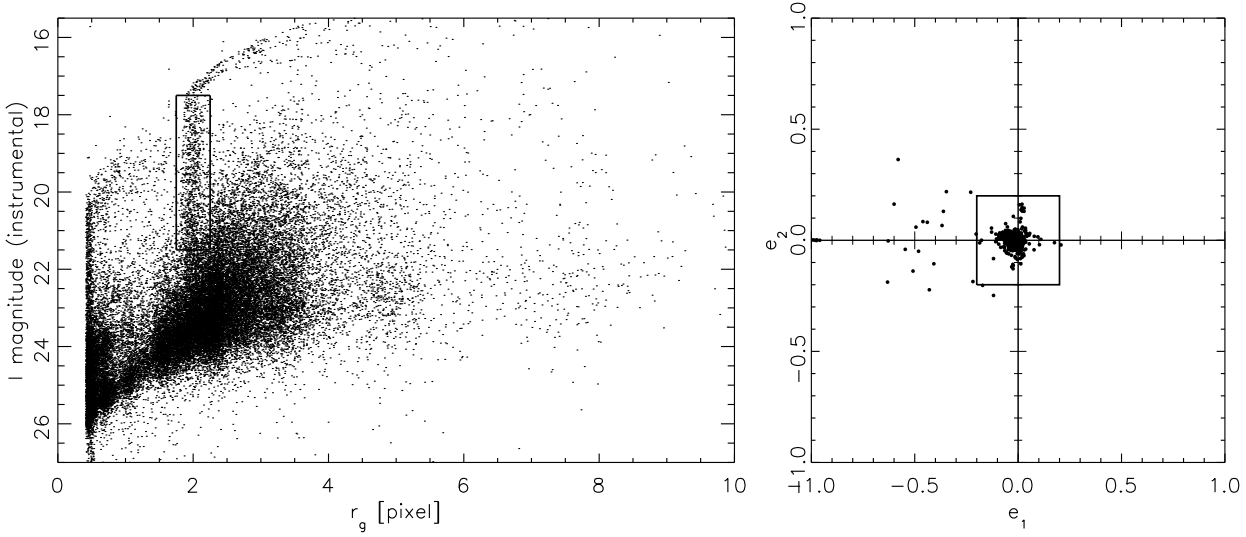}
  \caption{Example of size-magnitude (left) and weighted ellipticity
    parameter diagram (right) for an arbitrary exposure. The
    moderately bright, non-saturated stars are chosen from within the
    squares and utilised in the astrometric solution, the right plot
    containing only the stars chosen from the left plot. As square is
    the only option {\tt IMCAT} offers in selecting objects, this is
    also
    applied to the ellipticity diagram. Although a circle might yield
    better results, the difference is considered negligible. The stars
    utilised in the PSF corrections (Sect.~\ref{psfcorr}) are selected
    the same way.}
  \label{ast}
\end{figure*}

First, all objects in each exposure were detected and
aperture photometry carried out. By plotting $r_g$ ($\sim$half-light
radius) vs. instrumental magnitude of the objects in each exposure, we
extract the moderately bright,
non-saturated stars suitable for deriving an astrometric
solution. Their weighted ellipticity parameters \citep[defined
by][]{ksb95}, $e_{1,2}$, are in addition plotted and eyeballed,
selecting the main clustering of objects to ensure a catalogue
containing purely stars (see Fig.~\ref{ast} for an example).

Left with star catalogues we now compute the transformation parameters
needed using information from the USNO-B1.0 catalogue \citep{mlc03}.
However, as many of the USNO-B1.0 stars are saturated in our images,
we extended our reference catalogue by detecting more sources in
{\tt FITS} files derived from the Digitized Sky Survey
(DSS)\footnote{\tt http://archive.stsci.edu/dss/index.html}.
We match the target catalogues to the reference catalogue, then solve
for a set of low-order spatial polynomials mapping the images onto
each other, by repeatedly refining the least squares minimisations
using outlier rejection.

\subsubsection{Final master image and object catalogue}\label{final}

Gain and quantum efficiency variations between chips and differential
extinction between exposures are solved for by least squares
minimisations. The extinction corrections between the exposures were
very small, typically \mbox{$\sim 0.01$} mag, whereas the zero-point
offsets between the chips were \mbox{$\sim 0.1$} mag. As an accurate
absolute photometric calibration is not necessary for the present
work, we adopted standard Landolt magnitude zeropoints.

The co-addition is done after magnitude corrections are applied to the
data. In addition cosmic rays are masked out, before the median image
is computed and the background flattened. A master object catalogue is
created, where each object's WCS coordinates are calculated from the
astrometric solution, and their ellipticity parameters computed (see
Sect.~\ref{shear}). Finally we mask out false detections by
over-plotting the objects onto the image, hence ensuring a final
object catalogue free from spurious detections.

\section{Shear measurements}\label{shear}

Identifying weak lensing effects requires measuring the ellipticities
of a large number of faint background galaxies. The main source of
noise in weak lensing analysis is the intrinsic ellipticities of these
galaxies. To distinguish between distorted images resulting from a
weak lens and the usual distribution of shapes existing in an unlensed
galaxy population, the ellipticities are examined for a systematic
change. In particular, a tangential alignment of the galaxy shapes
around the cluster centre would confirm the existence of a weak lens.

An additional source of error comes from the faint background galaxies
being smeared by the PSF, caused by atmospheric turbulence and
optical aberrations. The weak
shear signal is hence diluted because this smearing will cause the
galaxy images to appear more circular than before the smearing.
In addition, PSF anisotropies distort the images, causing the galaxies
to appear more elliptical, hence introducing false shear signals.
It is crucial that these PSF effects are corrected for.

\subsection{PSF corrections}\label{psfcorr}

Following the KSB+ method developed by \citet{ksb95}, \citet{lk97},
and \citet{hfk98}, we present a short summary of our implementation
below. KSB+ inverts the effects of PSF smearing and anisotropy on
objects in an image, presenting a method to recover the true shear.

\begin{figure*}
  \includegraphics[width=17cm]{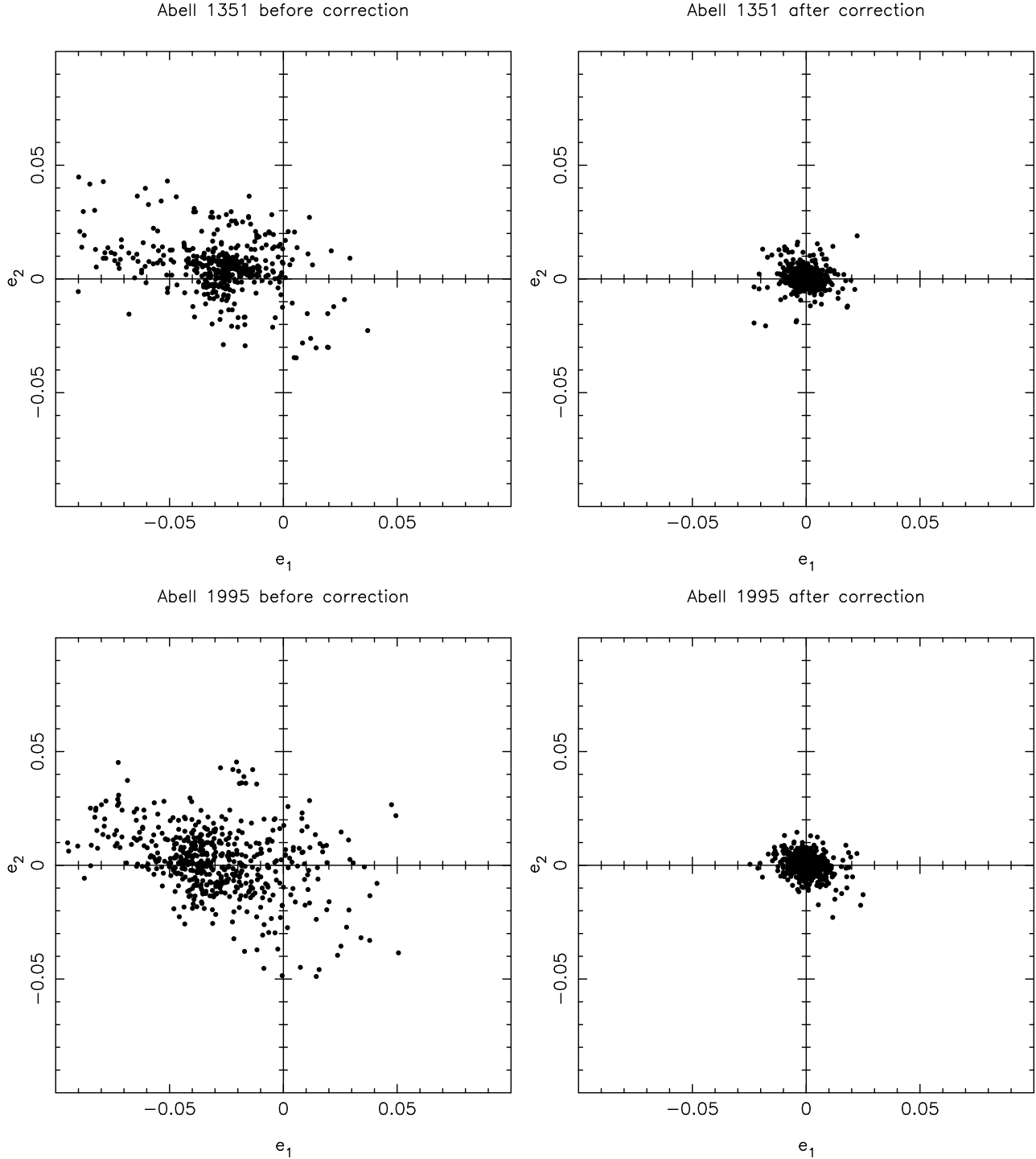}
  \caption{Ellipticities of the stars in the field of Abell~1351 (top)
    and Abell~1995 (bottom) before and after corrections for PSF
    anisotropies. The stars initially have systematic ellipticities up
    to $\sim7-9$\% in one direction. The PSF corrections reduce these
    effects to typically $<1.5$\%.}
  \label{starsall}
\end{figure*}

Ignoring the effects of photon noise, it is possible to
express the observed ellipticity of a galaxy as
\begin{equation} \label{eobs}
  e_\alpha^{\rm obs} = e_\alpha^{\rm s} + P_{\alpha\beta}^\gamma
  g_\beta + P_{\alpha\beta}^{\rm sm}p_\beta ,
\end{equation}
where the first term represents the intrinsic ellipticity of the
galaxy, the second term the shift in ellipticity caused by
gravitational shear, and the third term the smearing of the galaxy
image from the anisotropic PSF \citetext{see \citealp{lk97} with
  additional corrections from \citealp{hfk98} for a thorough deduction
  of this equation}. Present in eq.~(\ref{eobs}) are the pre-seeing
shear polarisability tensor, $P_{\alpha\beta}^\gamma$, and the smear
polarisability tensor, $P_{\alpha\beta}^{\rm sm}$. The latter is
calculated for each object together with $e_\alpha^{\rm obs}$, the
post-seeing shear polarisability tensor $P^{\rm sh}_{\alpha\beta}$,
and the centroid of the object.

Because stars are foreground objects ($g_\beta=0$) and intrinsically
circular ($e_\alpha^{\rm s}=0$), applying eq.~(\ref{eobs}) to stellar
objects provides a measure of the total PSF anisotropy,
$p_\beta$. This is calculated from
bright stars selected from the final object catalogue
(Sect.~\ref{final}). The PSF corrections are then
calculated for all individual objects and corrections applied
respectively.

The ellipticities of the stars were fitted to a sixth-order Taylor
series expansion. When comparing mass and B-mode maps
(Sect.~\ref{mass}) for fits of different orders, there was little
change with the order of fit. Over the whole field, 410 and 530 stars
were used in the fitting process for Abell~1351 and Abell~1995
respectively. Fig.~\ref{starsall} shows the ellipticities of the stars
before and after PSF corrections.

The pre-seeing shear polarisability tensor, $P^\gamma_{\alpha\beta}$,
is defined in KSB+ to be
\begin{equation}\label{pgammaalphabeta}
  P^\gamma_{\alpha\beta} = P^{\rm sh}_{\alpha\beta} - 
  P^{\rm sm}_{\alpha\mu}\; (P^{\rm sm\star})^{-1}_{\mu\delta}
  \; P^{\rm sh\star}_{\delta\beta} ,
\end{equation}
where the asterisk denotes $P^{\rm sh}_{\alpha\beta}$ and 
$P^{\rm sm}_{\alpha\delta}$ applied to stellar objects. From
eq.~(\ref{eobs}) we see that the reduced shear,
\mbox{$g=\gamma/(1-\kappa)$}, is given by
\begin{equation}\label{gammaalpha}
  g_\beta = (P^\gamma)^{-1}_{\alpha\beta} 
  \left[e_\alpha^{\rm obs} - P^{\rm sm}_{\alpha\beta}p_\beta\right]
  .
\end{equation}
Following the approach by \citet{wld02}, we assume the PSF is close to
circular after the correction, and the polarisabilities can be
approximated by
\mbox{$P_{\alpha\beta}=\frac{1}{2}(P_{11}+P_{22})\delta_{\alpha\beta}$}. The
average of \mbox{$P^{\rm sh\star}/P^{\rm sm\star}$} can be calculated
as
\begin{equation}
  \left< \frac{P^{{\rm sh}\star}}{P^{{\rm sm}\star}} \right> = 
  \frac{1}{N_{\rm stars}} \sum_{\rm stars} 
  \frac{P^{\rm sh\star}_{11}+P^{\rm sh\star}_{22}}{P^{\rm sm\star}_{11}+P^{\rm sm\star}_{22}} ,
\end{equation}
where we use the median value rather than the mean to minimise
the effect of outliers. Hence eq.~(\ref{pgammaalphabeta}) turns
into
\begin{equation}\label{pgamma}
  P^\gamma = \frac{1}{2} (P^{\rm sh}_{11}+P^{\rm sh}_{22}) -
  \frac{1}{2} (P^{\rm sm}_{11}+P^{\rm sm}_{22}) 
  \left<\frac{P^{{\rm sh}\star}}{P^{{\rm sm}\star}}\right> .
\end{equation}
As $P_{\alpha\beta}^{\rm sh}$ and $P_{\alpha\beta}^{\rm sm}$ are
already calculated, this equation is easily solved, and
eq.~(\ref{gammaalpha}) gives us an estimate of the gravitational shear
of each object. \citet{hfk98} show that estimating the PSF dilution
for each individual galaxy introduces additional noise. We therefore
follow their approach by determining $P^\gamma$ as a function of
magnitude and galaxy size. We determine the median $P^\gamma$ within
15 bins in an $r_g$-magnitude diagram, where the central 4 bins
contain $\sim4000$~galaxies/bin and the outer ones
$\sim200$~galaxies/bin. We then compute one correction factor for each
bin using eq.~(\ref{pgamma}), and apply this to all galaxies within
the corresponding bin.

The faintest and smallest galaxies are more affected by seeing than
the larger galaxies, giving them a poorer shape determination and a
larger correction factor. Such galaxies are therefore of less
importance. To account for this, a normalised weight,
\begin{equation} \label{weight}
  w_i \propto \left( \frac{\sigma_{e_i}}{\left<P^\gamma\right>_i}
  \right)^{-2} ,
\end{equation}
is calculated for each bin $i$ and assigned to the corresponding
galaxies.
Here, $\sigma_{e_i}$ is the observed dispersion of galaxy
ellipticities.

\section{Mass reconstruction}\label{mass}

We select background galaxies with $6<{\rm S/N}<100$ for the creation
of our mass maps. These are reconstructed from the galaxies' shapes
using the finite-field inversion method presented by \citet[][SS01]{ss01}. For
this method a smoothed shear field is calculated on a grid using a
modified Gaussian filter. The algorithm then iteratively computes a
quantity $K(\theta) :={\rm ln[1-\kappa(\theta)]}$ which is determined
up to an additive constant due to the mass sheet degeneracy. We break
the latter by assuming that the average convergence vanishes along the
border of the wide field of view. The width of the Gaussian term in
the filter is set to $2\farcm0$, resulting in an effective smoothing
length of about $1\farcm6$.

\begin{figure*}
  \resizebox{\hsize}{!}{\includegraphics{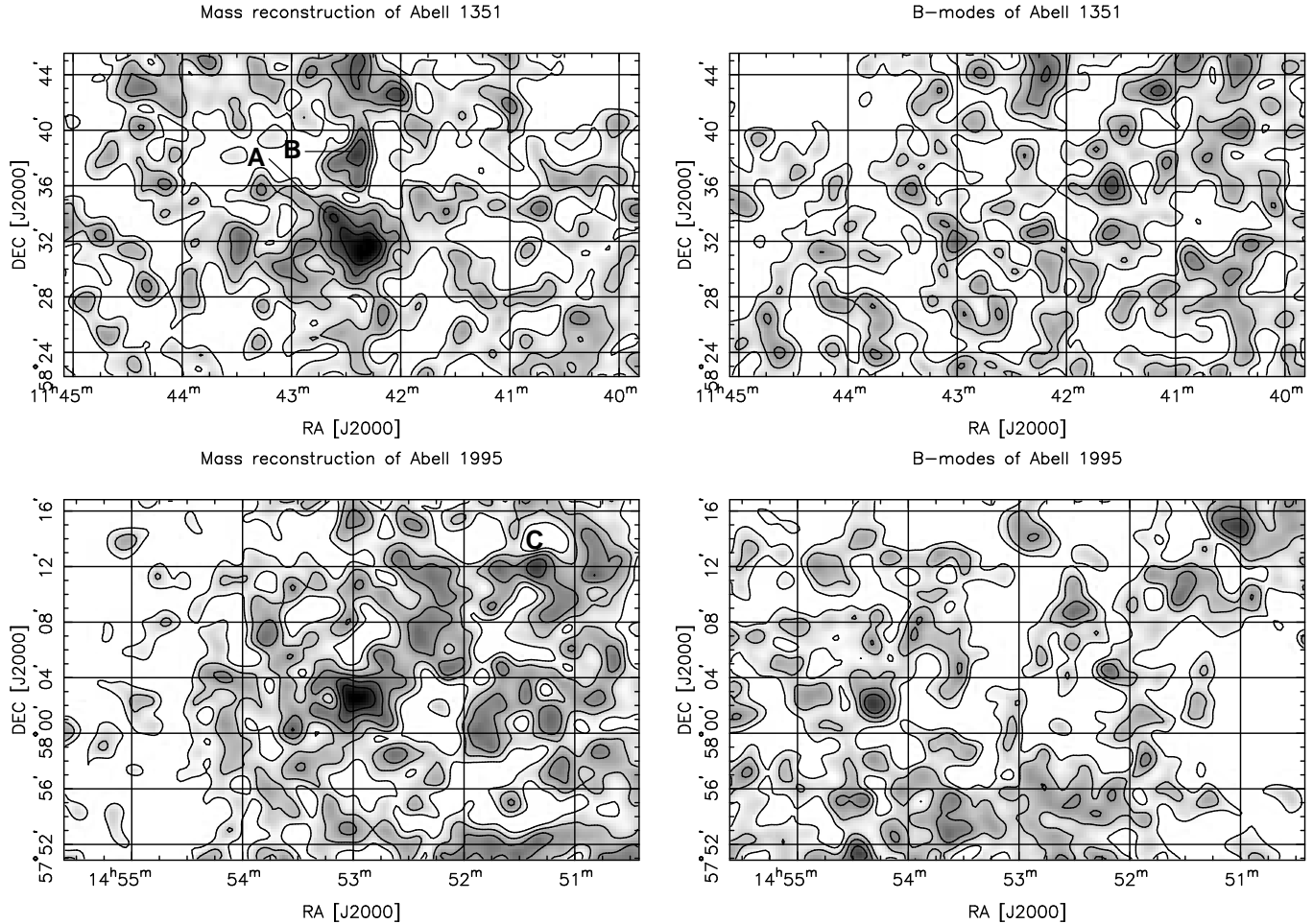}}
  \caption{The projected surface mass densities and B-modes for both clusters 
    in the full CFH12K field, using a finite-field mass
    reconstruction. The maps show the S/N of the clusters, with
    contours starting at $0\sigma$ and increasing in steps of
    $1\sigma$. The clusters are detected at significance levels of
    \mbox{$5.3\sigma$} (Abell~1351) and \mbox{$5.2\sigma$}
    (Abell~1995). The smaller peaks A, B, and C have S/N-ratios of
    4.2, 3.8, and 3.8, respectively. The effective smoothing length of
    the reconstructions is $1\farcm6$. The highest B-mode peak of
    Abell~1351 (Abell~1995) has a S/N-ratio of 3.5 (3.9).}
  \label{bothgrey}
\end{figure*}

In order to evaluate the noise of the mass maps, we computed $2000$
mass maps for each cluster based on randomised galaxy orientations,
keeping their
positions fixed. As the cluster lens signal increases the
ellipticities of galaxies, this would lead to an overestimation of the
noise at the cluster position. We roughly correct for this effect by
subtracting the expected SIS tangential shear signal, determined from
the clusters' known velocity dispersions (see
Table~\ref{results}). Since the singularity of the SIS can lead to
overly large corrections close to the cluster centre, we limited the
maximum correction factor allowed to 0.5 in each ellipticity
component. This affected less than 5 galaxies in both fields. The true
mass maps were then divided by the noise maps obtained from the
randomised mass maps to create the S/N-maps seen in
Fig.~\ref{bothgrey}.

Abell~1351 and Abell~1995 are detected with a S/N of $5.3$ and $5.2$,
respectively. Upon integrating the $\kappa$ maps within
$r_{200}=1.69\;h_{70}^{-1}\;{\rm Mpc}$ ($1.50\;h_{70}^{-1}\;{\rm Mpc}$) for
Abell~1351 (Abell~1995), we find total masses of
\mbox{$M_{\rm 2D}(<r_{200})=11.7\pm3.1\times10^{14}\;h_{70}^{-1}M_\odot$}
and
\mbox{$M_{\rm 2D}(<r_{200})=10.5\pm2.7\times10^{14}\;h_{70}^{-1}M_\odot$}
for the clusters, respectively. The $r_{200}$ radii are taken from
what we consider to
be the best NFW fits to the data (see Table~\ref{results} and
Sect.~\ref{sisnfw}), while the errors are determined from integrating
the same areas in the 2000 noise maps.

In Fig.~\ref{bothgrey} the B-modes in both cluster fields are also
shown, computed by repeating the $\kappa$ reconstruction with all
galaxies rotated $45\degr$ \citep{cnp02}. Provided that the lensing
data are free from systematics and that the noise (intrinsic 
ellipticities) is Gaussian, this B-mode map should be consistent with 
Gaussian noise. Given the effective filter scale of $1\farcm6$, 
we can place about 380 independent peaks in the CFH12K field. 
Thus one would expect $1.1$ noise peaks above $3\sigma$ in the field. 
A more realistic estimate comes from the 2000 randomisations as these
are based on the real ellipticity and spatial distribution. We 
expect $1.4$ ($1.6$) such peaks for Abell~1351 (Abell~1995). In the
real B-mode maps we find $3$ peaks for each of the clusters. This
is insignificant, as in our randomisations at least 3 such peaks
appear per field in $20\%$ of the cases. In case of
Abell~1995 the largest B-mode peak has a significance of $3.9\sigma$.
Its B-modes appear generally somewhat larger than for Abell~1351, 
which has no B-mode peaks higher than $3.5\sigma$. 

\subsection{Mass and galaxy density distributions}\label{verify}

In order to compare the surface density maps with the distribution of
cluster galaxies, we extract the red sequence \citep[see e.g.][]{gy00}
and investigate the distribution of the galaxies selected. We combine
the \emph{I}-band data with the \emph{V}-band images from
\citet{dki02}. To match the two data sets we re-sampled both images to
the same pixel scale, resulting in a common area of
\mbox{$18\farcm5\times18\farcm5$} on the sky. The \emph{V}-band image
seeing was around $0\farcs7$, and thus consistently better than in the
\emph{I}-band. The \emph{V}-band data were therefore convolved to
match the seeing in the \emph{I}-band data, $0\farcs95$ for
Abell~1351 and $1\farcs15$ for Abell~1995. Aperture photometry was
carried out using {\tt SExtractor} \citep{ba96} in double-image
mode. The deep \emph{I}-band images served as the detection image,
providing us with a target list with defined coordinates. At these
positions we integrated the flux in a $3\arcsec$ wide aperture in each
of the \emph{V}- and \emph{I}-band images. Plotting the galaxies in a
colour-magnitude diagram will then in principle provide enough
information to separate the red early-type cluster members from
the other galaxies.

\begin{figure*}
  \resizebox{\hsize}{!}{\includegraphics{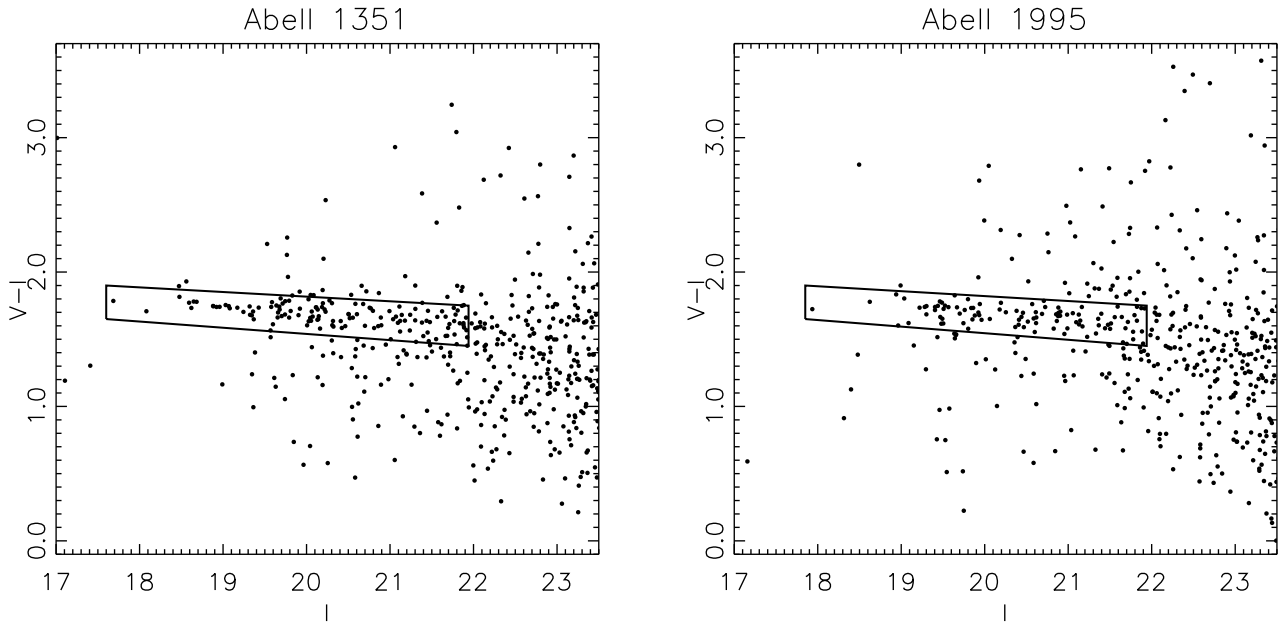}}
  \caption{Colour-magnitude diagrams for Abell~1351 (left) and
    Abell~1995 (right), where only galaxies within $3\arcmin$ from the
    BCG are plotted. The red sequence can be seen for
    \mbox{$1.4<V-I<1.9$} for both clusters, the box
    indicating our selection criteria.}
  \label{colmagdiag}
\end{figure*}

\begin{figure*}
  \resizebox{\hsize}{!}{\includegraphics{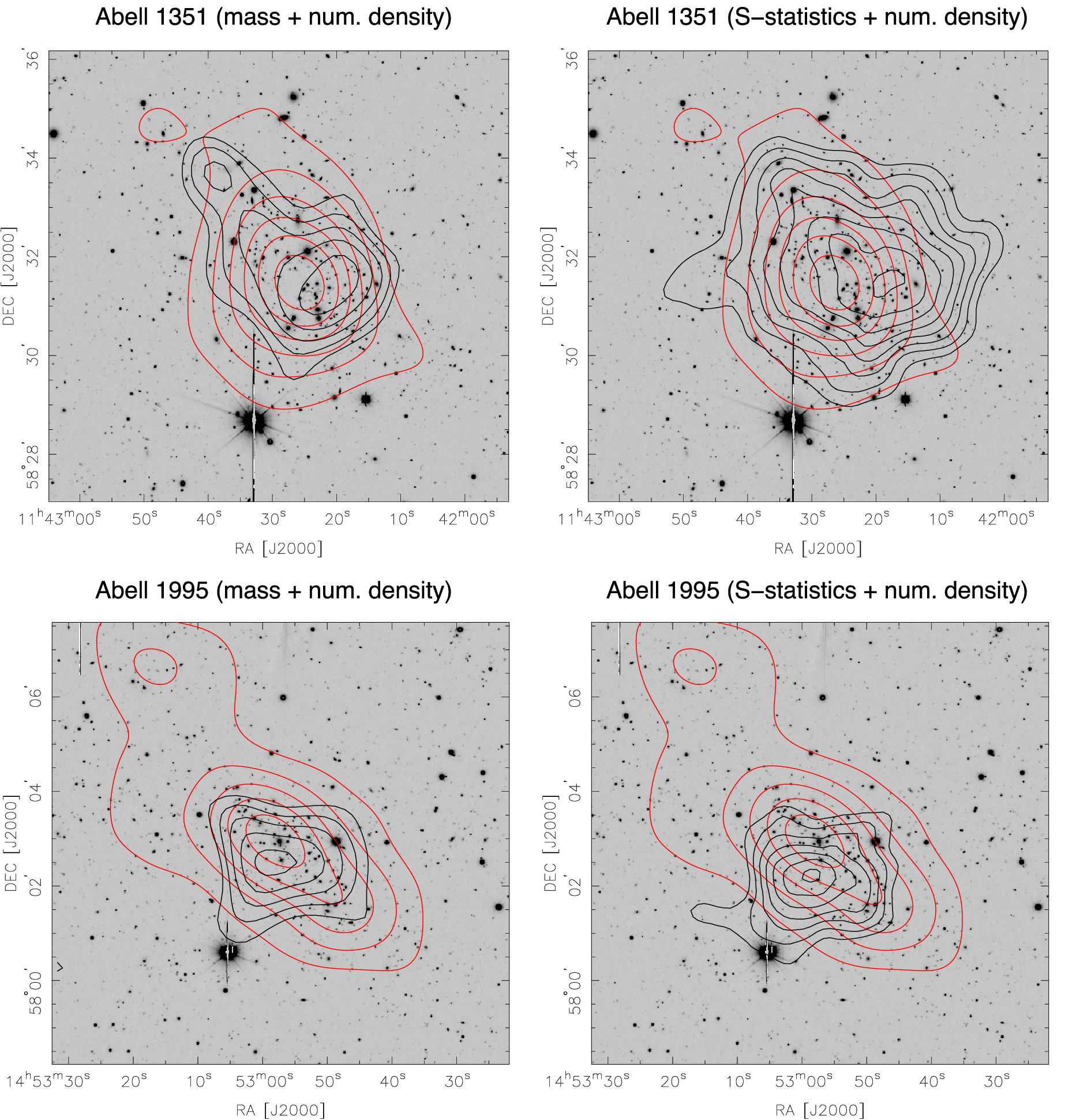}}
  \caption{The black contours show the mass reconstruction (left) and
    $S$-statistics (right) for Abell~1351 (top) and Abell~1995
    (bottom). The contours start at the $3.0\sigma$ level and increase
    in steps of $0.5\sigma$. Over-plotted in white contours (red
      in online version) are the
    number densities of galaxies selected with the red sequence
    method, normalised by the rms fluctuation in the field outside the
    clusters. These contours are isodensity contours, starting at
    $3\sigma$ going in steps of $3\sigma$, and peak at $15$ and
    $10\sigma$, respectively.}
  \label{galdens}
\end{figure*}

Each cluster's red sequence does not clearly stand out from the
\mbox{$V-I$} vs. \emph{I} diagram when all objects are plotted.
We therefore select only galaxies within a radius of $3\arcmin$ of the
brightest cluster galaxy (BCG) for the colour-magnitude diagram (see
Fig.~\ref{colmagdiag}), and detect the red sequence at
\mbox{$1.4<V-I<1.9$} for both clusters. The selection
criteria indicated by the box in each plot are then applied to the
entire object catalogue. The number density of the galaxies selected
is then calculated as a function of position and over-plotted
onto the central $9$~arcmin of the mass maps (see
Fig.~\ref{galdens}).

We normalise the number density maps by the fluctuation measured in
the field outside the clusters. The centres of Abell~1351 and
Abell~1995 are then detected with $15.5\sigma$ and $9.9\sigma$
significance, respectively. The positional offsets between mass
centres, BCGs, and centres of galaxy density distributions are in
the range of $30\arcsec - 55\arcsec$ for both clusters, and are
due to noise in the mass maps. Changing the width of the Gaussian
kernel in the finite-field reconstruction algorithm shows that the
peak centres can drift by up to $20\arcsec$ from the mean
position. These offsets are consistent with other results in the
literature, such as \citet{cbg06}, who observed offsets on the order
of $10\arcsec$ between the lensing peaks (of higher S/N than ours) and
the optical centres of the bullet cluster. Positional offsets of
$50\arcsec$ are common in the sample of $70$ shear-selected clusters
by \citet{seh07}.

The cluster galaxy distribution resembles well the mass distribution
in the central part of Abell~1351. It extends significantly towards
the north-east, a feature also seen in the mass map where we find a
local maximum which we refer to as peak A (see
Sect.~\ref{substructure}). The galaxy distribution of Abell~1995
appears elliptical and elongated in the northeast-southwest
direction. This property is not reflected in the mass map where the
peak is of rather circular appearance.

To check the integrity of our mass reconstructions further, we applied the
peak finder ($S$-statistics) developed by \citet{seh07}. This method 
detects areas of enhanced tangential shear using the aperture mass statistics
\citep{sch96}. Since it uses a filter function that mimics the tangential 
shear profile of galaxy clusters it is well suited for detecting mass
concentrations. With this method we recover Abell~1351 at the $7.0\sigma$
level in a $10\arcmin$ wide filter, and Abell~1995 with $6.1\sigma$ for a
$7\arcmin$ filter. The filter shape parameter \citep{seh07} was chosen as
$x_{\rm c}=0.2$ in both cases. We find that the $S$-statistics is elongated in
the same way as the mass reconstruction for Abell~1351, extending towards peak
A. We evaluate the significance of this possible sub-structure in
the following.

\subsection{Smaller mass peaks in the fields}\label{substructure}

In the mass reconstructions two neighbouring peaks A and B are
detected around Abell~1351, and another one (peak C) in the field of
Abell~1995. Their S/N-ratios are $4.2$, $3.8$, and $3.8$,
respectively. Using the 2000 noise randomisations for each field we
find that the probability of a noise peak larger than $4.2\sigma$
($3.8\sigma$) in the field of Abell~1351 is $0.8\%$ ($6.8\%$),
respectively. The corresponding probability for peak C in the
Abell~1995 data is $7.0\%$. These are somewhat larger
than what would be expected from purely Gaussian noise.

Hence the only significant sub-structure we detect in the
mass reconstruction is peak A near Abell~1351. Looking at the
contours in the upper right panel of Fig.~\ref{galdens} one can see
that the $S$-statistics trace this structure as well at the
$4.5-5.0\sigma$-level. We note that we recover this sub-structure over
a broad range of filter scales in the $S$-statistics and hence think
that it is a real feature in the mass distribution of Abell~1351.

Out of the broad range ($1\arcmin-15\arcmin$) of filters probed
with the $S$-statistics, peak B is detected only once with
$4.0\sigma$ in the $4\arcmin$ wide filter for $x_{\rm c}=0.5$. It has
the typical characteristics of the dark peaks found by \citet{seh07},
i.e. it is not associated with any over-density of galaxies. Hence it
is most likely a noise peak.

In the Abell~1995 field we could not detect any other peaks
using the $S$-statistics. Since the B-modes for those data show
a maximum of $3.9\sigma$ near peak C (at $3.8\sigma$), we
consider it to be a noise peak. As it also lies outside the area
covered by the \emph{V}-band, we could not check for over-densities of
red galaxies at this position.

\section{Modelling the lensing data}\label{model}

Comparing observed distortions in the background galaxies to those
predicted by theoretical density profiles enables us to estimate the
mass of a galaxy cluster. Using $\chi^2$-minimisations of SIS and NFW
models, we first determine the best fit parameters and then calculate
the cluster masses.

The theoretical profiles are both spherically symmetric. We therefore
average the tangential reduced shear,
\mbox{$g_{\rm t}=\gamma_{\rm t}/(1-\kappa)$} (for $r>\theta_{\rm E}$,
where $\theta_{\rm E}$ is the Einstein radius), in 17 radial bins
around the cluster centre. The bins are logarithmically spaced,
covering the entire field of view, and starting at
$r_{\rm min}=150\arcsec$ to avoid the large contamination from cluster
galaxies close to the centre of the field (see also
Sect.~\ref{clustcont}). To determine the cluster centre, we
tested three different positions.
First we adopt the peak location
in the mass reconstructions generated (Sect.~\ref{mass}). These
coincide with the centres of the $S$-statistics. Second, the BCG
serves as cluster centre, and third we try the centre of the galaxy
density of each cluster. As the latter coincide with the BCG for
Abell~1995, only two positions were tested for this cluster.
We also considered strong lensing features, but found that they 
do not offer further insight in this respect (see
Sect.~\ref{stronglensfeatures} for details).
We calculate $\left<g_{\rm t}\right>_i$ for each radial bin $i$
  and compare them to the theoretical values at
the average radius of each bin, $\left<r\right>_i$.

When calculating the mass of a cluster, the relative distance of the
background galaxies and the lensing cluster is required. As we have no
specific information about the redshifts of the background galaxies,
the distances have to be estimated statistically. By using the
photometric redshift distribution of corresponding faint galaxies from
the Hubble Deep Field (HDF) North \citep{fly99}, we can estimate the
average \mbox{$\beta\equiv D_{\rm ds}/D_{\rm s}$}, where $D_{\rm ds}$
is the angular diameter distance between the lens and the source and
$D_s$ between the observer and the source. The empirical relation
\begin{equation}
  \left<\beta\right> = -1.21 z_{\rm d} + 0.91
\end{equation}
is derived for a $\Lambda$CDM cosmology analogously to the equation of
\citet{dki02} for an Einstein-de~Sitter universe, which also accounts
for the redshift distribution of the source galaxies. Here
  $z_{\rm d}$ denotes the redshift of the lens.

\subsection{Cluster contamination and magnification
  depletion}\label{clustcont}

At small projected radii from the cluster centre our faint galaxy
sample will contain cluster galaxies in addition to background
galaxies. We could not discriminate faint cluster members from
lensed field galaxies using \mbox{$V-I$} colours, hence the sample of
presumed lensed background galaxies remains contaminated.
This leads to a systematic bias of the shear measurements towards
smaller values. 

\begin{figure}
  \resizebox{\hsize}{!}{\includegraphics{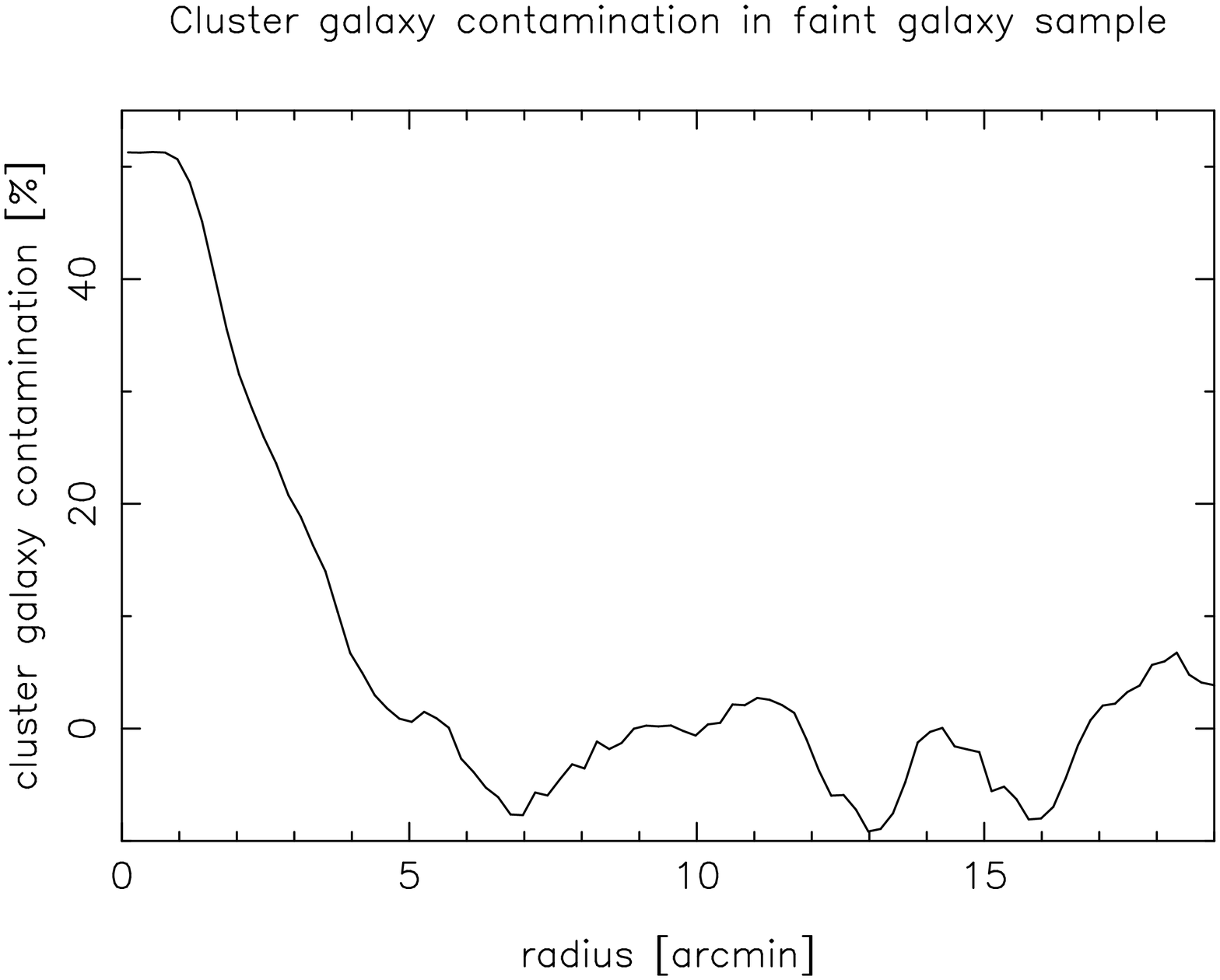}}
  \caption{Percentage of cluster galaxies in the faint galaxy
    catalogue of Abell~1351 (that of Abell~1995 is very
    similar). Because the projected density of cluster
    galaxies is assumed to equal zero at the edge of the field, the
    cluster galaxy contamination is set to zero here by subtracting
    the median value outside the central area of the image (the
    large field of view makes this a well-working approximation). Our
    fitting procedure starts at $r_{\rm min}=150\arcsec$ to avoid the
    large cluster galaxy contamination at the centre.}
  \label{clusgalcont}
\end{figure}

In order to quantify this contamination, we determine the
over-density of galaxies in the background catalogue at the cluster
position compared to the mean density in the field (an example
  is shown in
Fig.~\ref{clusgalcont} for Abell~1351). A contamination rate of $50\%$
is found for both cluster centres, vanishing for radii larger
than about $4\arcmin-5\arcmin$. It is in fact even higher, as
magnification depletion leads to a reduced number density of
lensed galaxies in the \emph{I}-band near the cluster centre.
However, this effect can be neglected in our case. From
the smoothed convergence (see Sect.~\ref{mass}) and reduced shear
fields we estimate the magnification using
$\mu=((1-\kappa)(1-g))^{-2}$. We find very similar magnifications
for both clusters, being $1.65$ at the centre and becoming
indistinguishable from the noise ($\sigma_\mu\sim0.15$) for radii
larger than $\sim3\arcmin$. The depletion of the number density 
of lensed galaxies is $\propto\mu^{2.5s-1}$, with $s=0.15$ in red
filters \citep[see e.g.][]{nb96}. At the cluster centres the number
densities are thus reduced by a factor of $\sim0.73$, and at a radius
of $1\farcm5$ magnification depletion becomes indistinguishable from
the natural
fluctuations in the distribution of field galaxies. Magnification
depletion hence only affects the innermost $\sim0.3\;{\rm Mpc}$
($100\arcsec$) of the clusters and can be neglected
since we compare the tangential shear profiles to models only for
radii larger than $0.5\;{\rm Mpc}$ (see Fig.~\ref{gtvsradius}).

In order to correct for the contamination by cluster galaxies,
we modify the theoretical shear values.
The reason for adjusting the theoretical values rather than 
the measured values is that this method is considerably easier to 
implement. The correction factors are determined in radial bins of
logarithmic spacing. One correction value is then calculated for each 
of the 17 bins in which \mbox{$\left<g_{\rm t}\right>$} is measured.
By assuming the edges of the field to 
be approximately free from cluster galaxies, the outermost correction 
factor can be set to 1 to mimic contamination-free boundaries of the 
field. Finally the best fit is found using $\chi^2$-minimisations.

\subsection{Fitting the SIS and the NFW profiles}\label{sisnfw}

Once the cluster centre is determined, the only free parameter of the
SIS profile is the velocity dispersion, $\sigma_v$. The best fit of
the SIS profile is determined by $\chi^2$-minimisation for a range of
$\sigma_v$ values, the results being shown in Table~\ref{results}. The
mass estimate, $M_{\rm SIS}$, for this profile is calculated at
$r_{200}$ (the radius inside which the mean mass density of the
cluster equals $200\rho_{\rm crit}$) found in the NFW fitting with two
free parameters utilising the same cluster centre.

The NFW profile is derived from fitting the density profiles of
numerically simulated cold dark matter halos. It appears to give a
very good description of the radial mass distribution inside the
virial radius of a galaxy cluster. For a thorough introduction to the
gravitational lensing properties of the NFW mass density profile
we refer the reader to \citet{wb00}. The theoretical $\gamma_{\rm t}$
and $\kappa$ can be calculated analytically for the NFW density
profile \citep{bs01}. We derive the best fit parameters for different
values of the concentration parameter, $c$, ranging from $0.1$ to
$24.9$ in steps of $0.1$.

With the cluster centre fixed, the NFW profile has two free
parameters, $r_{200}$ and $c$. We fitted our shear measurements to
this profile twice; first by keeping $c$ fixed and varying only
$r_{200}$ to find our best fit, and second by varying both
parameters. The best fit parameters were determined by minimising
$\chi^2$ in both cases. Based on N-body simulations of dark matter
halos, \citet{bks01} derive relations for the mean value of $c$ as a
function of redshift and mass for different cosmologies. For a halo of
mass \mbox{$M_{\rm vir}=8\times10^{14}M_\odot$}, the relation yields
\begin{equation}\label{bestfitc}
  c = \frac{5.8}{1.194(1+z_{\rm d})}
\end{equation} 
(where \mbox{$r_{200}=1.194r_{\rm vir}$} for a \mbox{$\Lambda$CDM}
cosmological model). As this mass is close to the mass estimates of
Abell~1351 and Abell~1995 (Sect.~\ref{mass}), the weak mass dependence
of $c$ can be
disregarded. In the second fitting process both $r_{200}$ and $c$ were
altered, creating a grid of \mbox{$c, r_{200}$}-values. The best fit
$r_{200}$ was first determined for each value of $c$, then the best
fit $c$ was found. The results are given in Table~\ref{results}.
We could not find an upper limit for the concentration parameter of
Abell~1351, independent of the cluster centre chosen. The same holds
for Abell~1995 in case the BCG is chosen as the centre.
We discuss this in more detail in Sect.~\ref{concparam}.

\begin{figure*}
  \resizebox{\hsize}{!}{\includegraphics{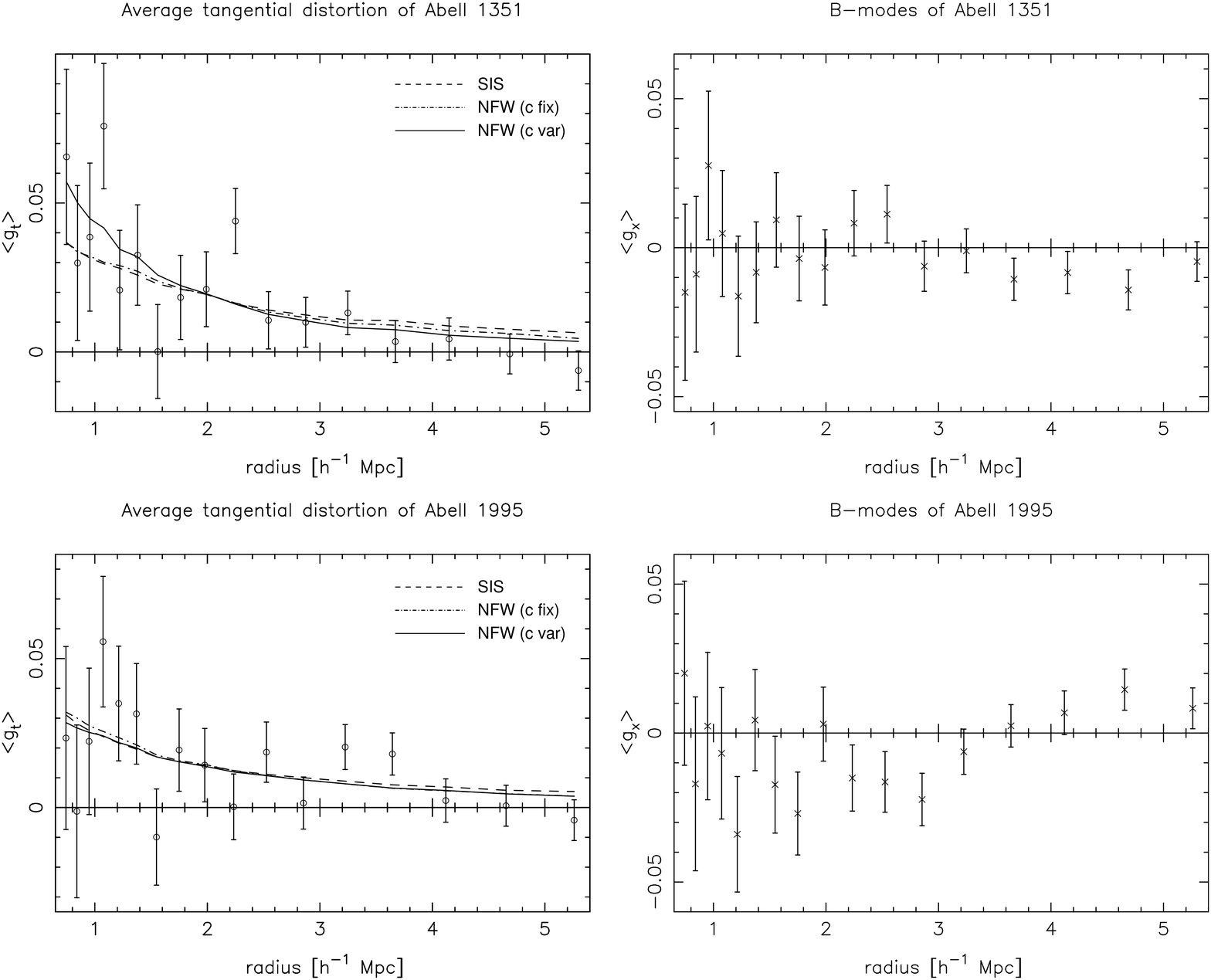}}
  \caption{\emph{Left:} Reduced tangential shear as a function of
    radius for Abell~1351 (top) and Abell~1995 (bottom) using the SS01
    $\kappa$ map as cluster centre (the other centres show very
    similar figures). The averaged gravitational lensing distortions
    of background galaxies are shown as points with $1\sigma$ error
    bars. The lines indicate the best fit models; the dashed line
    representing the SIS profile, the dot-dashed line the NFW profile
    keeping $c$ fixed, and the solid line the NFW profile with two
    free parameters. It should be noted that these lines represent the
    theoretical values \emph{after} modifications from cluster galaxy
    contamination are applied to each bin independently of each other
    (see Sect.~\ref{model}). \emph{Right:} Cross-component of the
    reduced tangential shear as a function of radius for Abell~1351
    (top) and Abell~1995 (bottom). This signal should disappear if
    $\left<g_{\rm t}\right>$ is caused by lensing only, and it is seen
    that the measurements are close to zero for both clusters.
  }
  \label{gtvsradius}
\end{figure*}

As an example we display the reduced tangential shear as a function of
radius using the SS01 $\kappa$ maps as cluster centre, see the left
diagrams of Fig.~\ref{gtvsradius}. The measured values of
$\left<g_{\rm t}\right>$ are given together with the best fit models
of the SIS and NFW profiles. Judging from the diagrams alone, the NFW
profile letting both $c$ and $r_{200}$ vary seems to represent the
best fit to the clusters. However, the $\chi^2/{\rm Dof}$ values given
for each model in Table~\ref{results} show that the differences
between the models are not statistically significant. The differences
emerging from different cluster centres seem to have a larger
impact. The right diagrams of Fig.~\ref{gtvsradius} show the B-modes
of both clusters, i.e. the cross-component of $\left<g_{\rm t}\right>$
as a function of radius. Both measurements are consistent with zero.

\begin{table*}
  \caption{Results from fitting \mbox{theoretical} density profiles to
    the measured shear values, where each sub-headline
    indicates the cluster centre around which the fitting was
    done. The mass estimates for the SIS profile ($M_{\rm SIS}$)
    are calculated at the best fit $r_{200}$ from the NFW profile with
    two free parameters using the same cluster centre. The NFW profile
    denoted \emph{``fixed'' $c$} refers to the fitting process keeping
    $c$ fixed while varying only $r_{200}$, where $c$ is calculated
    from eq.~(\ref{bestfitc}) \citep{bks01}.
    In the
    bottom we add the results from integrating the $\kappa$ maps
    within our best fit $r_{200}$ values. Note that the KSB+ PSF
      correction tends to underestimate the shear by $10-15\%$, which
      in turn reduces the cluster masses up to $20\%$.}
  \label{results}
  \begin{minipage}[t]{\columnwidth}
  \centering
  \renewcommand{\footnoterule}{}
  \begin{tabular}{lcc|lcc}
    \hline \hline
    {\bf BCG}& {\bf Abell~1351}& {\bf Abell~1995}& {\bf
      SS01\footnote{\citet{ss01}} $\kappa$ map}& {\bf Abell~1351}&
    {\bf Abell~1995}\\ \hline
    {\bf SIS} [$\chi^2/\;{\rm Dof}$] &$1.60$ &$1.70$& {\bf SIS}
    [$\chi^2/\;{\rm Dof}$] &$1.44$ &$1.14$\\
    {$\theta_{\rm E}$ [$\arcsec$]}& $16.2{+2.3\atop-2.4}$&
    $12.2\pm2.6$& {$\theta_{\rm E}$ [$\arcsec$]}&
    $15.0{+2.5\atop-2.3}$&
    $12.0{+2.5\atop-2.8}$\\
    {$\sigma_v$ [${\rm km\;s}^{-1}$]}& $1040{+70\atop-80}$&
    $900{+90\atop-100}$&
    {$\sigma_v$ [${\rm km\;s}^{-1}$]}& $1000\pm80$&
    $890{+90\atop-110}$\\
    {$M_{\rm SIS}$ [$10^{14}h_{70}^{-1}M_\odot$]}& $8.7{+1.3\atop-1.4}$&
    $5.8{+1.2\atop-1.3}$& {$M_{\rm SIS}$ [$10^{14}h_{70}^{-1}M_\odot$]}&
    $7.8{+1.3\atop-1.4}$& $5.5{+1.3\atop-1.4}$\\
    \hline
    {\bf NFW} [$\chi^2/\;{\rm Dof}$] &$1.32$ &$1.64$& {\bf NFW}
    [$\chi^2/\;{\rm Dof}$] &$1.12$ &$1.11$\\
    $c$& $9.1{+\infty\atop-4.6}$& $5.2{+\infty\atop-3.4}$& $c$&
    $16.1{+\infty\atop-8.4}$& $3.0{+9.0\atop-1.9}$\\
    {$r_{200}$ [$h_{70}^{-1}\;{\rm Mpc}$]}&$1.73{+0.13\atop-0.10}$&
    $1.53{+0.17\atop-0.13}$&
    {$r_{200}$ [$h_{70}^{-1}\;{\rm Mpc}$]}&$1.69{+0.10\atop-0.11}$&
    $1.50{+0.20\atop-0.11}$\\
    {$M_{200}$ [$10^{14}h_{70}^{-1}M_\odot$]}& $8.1{+1.8\atop-1.4}$&
    $5.6{+1.9\atop-1.4}$& {$M_{200}$ [$10^{14}h_{70}^{-1}M_\odot$]}&
    $7.5{+1.3\atop-1.5}$& $5.3{+2.1\atop-1.2}$\\
    \hline
    {\bf NFW} (``fixed'' $c$) [$\chi^2/\;{\rm Dof}$] &$1.42$ &$1.66$&
    {\bf NFW} (``fixed'' $c$) [$\chi^2/\;{\rm Dof}$] &$1.31$ &$1.12$\\
    $c$& $3.7$& $3.7$& $c$& $3.7$& $3.7$\\
    {$r_{200}$ [$h_{70}^{-1}\;{\rm Mpc}$]}& $1.81{+0.057\atop-0.16}$&
    $1.56{+0.13\atop-0.16}$&
    {$r_{200}$ [$h_{70}^{-1}\;{\rm Mpc}$]}& $1.67{+0.16\atop-0.086}$&
    $1.51{+0.16\atop-0.14}$\\
    {$M_{200}$ [$10^{14}h_{70}^{-1}M_\odot$]}& $9.4{+0.9\atop-2.4}$&
    $5.9{+1.5\atop-1.8}$& {$M_{200}$ [$10^{14}h_{70}^{-1}M_\odot$]}&
    $7.4{+2.1\atop-1.1}$& $5.4{+1.7\atop-1.5}$\\
    \hline \hline
    {\bf KS93\footnote{\citet{ks93}} $\kappa$ map}&&& {\bf Galaxy
      density}&&\\ \hline
    {\bf SIS} [$\chi^2/\;{\rm Dof}$] &$1.30$ &$2.32$& {\bf SIS}
    [$\chi^2/\;{\rm Dof}$]& $1.17$&\\
    {$\theta_{\rm E}$ [$\arcsec$]}& $16.5{+2.6\atop-2.4}$&
    $10.4{+2.7\atop-2.6}$& {$\theta_{\rm E}$ [$\arcsec$]}&
    $16.5{+2.3\atop-2.4}$&\\
    {$\sigma_v$ [${\rm km\;s}^{-1}$]}& $1050\pm80$& $830{+100\atop-110}$&
    {$\sigma_v$ [${\rm km\;s}^{-1}$]}& $1050{+70\atop-80}$&\\
    {$M_{\rm SIS}$ [$10^{14}h_{70}^{-1}M_\odot$]}& $8.9\pm1.5$&
    $4.9{+1.3\atop-1.4}$& {$M_{\rm SIS}$ [$10^{14}h_{70}^{-1}M_\odot$]}&
    $9.0{+1.3\atop-1.5}$&\\
    \hline
    {\bf NFW} [$\chi^2/\;{\rm Dof}$]& $0.96$& $2.22$& {\bf NFW}
    [$\chi^2/\;{\rm Dof}$]& $0.88$&\\
    $c$& $11.2{+\infty\atop-4.9}$& $0.9{+2.1\atop-0.4}$& $c$&
    $11.3{+\infty\atop-6.7}$&\\
    {$r_{200}$ [$h_{70}^{-1}\;{\rm Mpc}$]}& $1.74{+0.11\atop-0.10}$&
    $1.50{+0.17\atop-0.24}$&
    {$r_{200}$ [$h_{70}^{-1}\;{\rm Mpc}$]}& $1.76{+0.10\atop-0.11}$&\\
    {$M_{200}$ [$10^{14}h_{70}^{-1}M_\odot$]}& $8.3{+1.6\atop-1.4}$&
    $5.3{+1.8\atop-2.6}$& {$M_{200}$ [$10^{14}h_{70}^{-1}M_\odot$]}&
    $8.5{+1.5\atop-1.7}$&\\
    \hline
    {\bf NFW} (``fixed'' $c$) [$\chi^2/\;{\rm Dof}$]& $1.13$& $2.33$&
    {\bf NFW} (``fixed'' $c$) [$\chi^2/\;{\rm Dof}$]& $1.00$& \\
    $c$& $3.7$& $3.7$& $c$& $3.7$& \\
    {$r_{200}$ [$h_{70}^{-1}\;{\rm Mpc}$]}& $1.81{+0.10\atop-0.13}$&
    $1.51{+0.11\atop-0.23}$&
    {$r_{200}$ [$h_{70}^{-1}\;{\rm Mpc}$]}& $1.79{+0.10\atop-0.11}$&\\
    {$M_{200}$ [$10^{14}h_{70}^{-1}M_\odot$]}& $9.4{+1.6\atop-2.0}$&
    $5.4{+1.2\atop-2.5}$& {$M_{200}$ [$10^{14}h_{70}^{-1}M_\odot$]}&
    $9.0{+1.5\atop-1.7}$&\\
    \hline \hline
    {\bf Integration of SS01 $\kappa$ maps}&&&&& \\ \hline
    {$r_{200}$ [$h_{70}^{-1}\;{\rm Mpc}$]}& $1.69$& $1.50$&&& \\
    {$M_{\rm 2D}(<r_{200})$ [$10^{14}h_{70}^{-1}M_\odot$]}&
    $11.7\pm3.1$& $10.5\pm2.7$&&& \\
    \hline
  \end{tabular}
\end{minipage}
\end{table*}

\section{Discussion}\label{discussion}

X-ray studies show Abell~1351 to be a galaxy cluster exhibiting
significant dynamical activity and undergoing a major merger event
\citep{asf03}, which indicates a cluster still in its formation
phase. Analyses assuming a relaxed cluster will hence naturally differ
from weak lensing analyses, where no assumption is made about the
dynamical state of the cluster. One example is the virial analysis by
\citet{ild02}, where an unusually high velocity dispersion of
\mbox{$\sigma_v=1680{+340\atop -229}$ km s$^{-1}$} is obtained for
Abell~1351, based on radial velocity measurements of 17 cluster
galaxies. Such a high velocity dispersion is not uncommon in merging 
systems. If for example two smaller clusters with low velocity
dispersions fall towards each other along the line of sight with
a velocity comparable to or larger than their $\sigma_v$, then
a very large total $\sigma_v$ would be inferred, with a 
correspondingly overestimated virial mass. The cluster CL0056.03-37.55 
is a good example for such a system \citep[see][]{ses03}.

Abell~1995 is, unlike Abell~1351, classified as a relaxed cluster in
dynamical equilibrium \citep{pd06}. X-ray studies and virial analyses
of this cluster are hence also more compatible with lensing studies
\citep{pjc00,ild02}. The projected two-dimensional distribution of
cluster galaxies in Abell~1995 is clearly elliptical (see
Fig.~\ref{galdens}), whereas the central lensing mass distribution is
circular.

\subsection{The mass estimates}

The mass distributions of Abell~1351 and Abell~1995 are estimated
assuming that the clusters follow spherically symmetric SIS or NFW
profiles. Although an elliptical mass profile might yield more
accurate cluster mass estimations, \citet{dsc05} experience that the
results from fitting a singular isothermal ellipse (SIE) model depend
strongly on the initial values chosen for the minimisation routines. 
We therefore decided not to fit the SIE profile to our clusters.

\citet{stepI} demonstrate in the Shear TEsting Program (STEP) that the
KSB+ PSF correction tends to systematically underestimate the shear
values \mbox{$\sim10-15\%$}. To measure how much this affects our
data, we calculate an upper limit for our mass estimates by increasing
the ellipticities with $15\%$ and repeating the fitting process. We
find that the underestimation of shear leads to an underestimation of
the total cluster mass with a maximum of
$20\%$, which is within the initial error bars.
The concentration parameters do not change significantly by
this boosting of ellipticities. Since we do not know by exactly
how much our shear values are underestimated this was merely an
attempt to quantify this effect on our data, and is not taken into
account in the results presented in this paper.

\citet{dki02} obtain weak lensing estimates of the cluster velocity
dispersion of several clusters using an SIS model and assuming an
Einstein-de~Sitter Universe. Their results are given for Abell~1351 as 
\mbox{$\sigma_v=1410{+80\atop -90}\;{\rm km\;s}^{-1}$} and for
Abell~1995 as \mbox{$\sigma_v=1240\pm80\;{\rm km\;s}^{-1}$}, and do
not agree with our results.
However, there are several important differences in methodology
between \citet{dki02} and our work. As mentioned above, the
assumed cosmological model is different. Also, these authors
approximate \mbox{$g_{\rm t}=\gamma_{\rm t}$}, whereas we use
\mbox{$g_{\rm t}=\gamma_{\rm t}/(1-\kappa)$} in our fits and mass
reconstructions.
Finally, the shear estimator of \citet{kai00} used by \citet{dki02}
is shown by \citet{stepI} to have a non-linear response to shear. A
re-analysis of the \citet{dki02} data, taking all these effects into
account, yields new values of
\mbox{$\sigma_v=1410\pm90\;{\rm km\;s}^{-1}$} and
\mbox{$\sigma_v=1000\pm100\;{\rm km\;s}^{-1}$} for Abell~1351 and
Abell~1995, respectively. Hence there still remains a systematic
discrepancy between the results for Abell~1351, while the measurements
for Abell~1995 agree within error bars.

A remaining difference between our work and \citet{dki02} is the
maximum radius,
$r_{\rm max}$, to which the shear is measured, given by the field of
view of the detector. Changing $r_{\rm max}$ in our Abell~1351 data to
$550\arcsec$ \citetext{as this is the value used by \citealp{dki02}}
yields \mbox{$\sigma_v=1240\pm105\;{\rm km\;s}^{-1}$}, consistent with
the re-analysed \citet{dki02} values within error bars.

\citet{asf03} use the \citet{dki02} observations to obtain a weak
lensing mass estimate applying the NFW model to a \mbox{$\Lambda$CDM}
cosmology. Their results give
\mbox{$M_{200}=30.2{+5.6\atop-4.9}\times10^{14}\;h_{70}^{-1}M_\odot$} for
Abell~1351 and
\mbox{$M_{200}=14.4{+3.3\atop-3.0}\times10^{14}\;h_{70}^{-1}M_\odot$} for
Abell~1995. These values are high compared to the results of this
study. \citet{asf03} use a fixed concentration parameter in the
fitting process, $c=5$. Applying this value to our data
yields minimal changes in $M_{200}$. The discrepancies hence originate
from \citet{asf03} utilising
\mbox{$r_{200}=2.69{+0.14\atop-0.19}\;h_{70}^{-1}{\rm Mpc}$}
and \mbox{$r_{200}=2.07{+0.19\atop-0.14}\;h_{70}^{-1}{\rm Mpc}$}
(priv. comm.) for Abell~1351 and Abell~1995, respectively, as these
values are larger than the $r_{200}$ values we obtain as our best
fit.

\subsection{The concentration parameter}\label{concparam}

The mass density of a cluster with a small concentration parameter
decreases slower when going to larger radii than for a cluster with a
large $c$ value \citep{wb00}. Although unconstrained upwards, we
find a lower limit of $c\geq4.5$ for Abell~1351.
As is also seen from the radial dependence of the shear in
Fig.~\ref{gtvsradius} (top left), the mass distribution of Abell~1351
concentrates around the cluster centre, indicating that its
concentration parameter is significantly higher than that of
Abell~1995. The found values for $c$ of Abell~1995 (see Table~\ref{results})
suggest that its mass is spread more evenly to larger radii, which is
also seen in Fig.~\ref{gtvsradius} (bottom left).

From their aperture mass calculations \citet{dki02} find that most of
the mass of Abell~1995 is contained within 
\mbox{$r\sim0.9\; h_{70}^{-1}{\rm Mpc}$} ($\sim200\arcsec$). The mass of
Abell~1351 shows the opposite behaviour, increasing evenly with radius,
even at large radii. These results are contrary to our conclusions.
As measurements at large radii are certain to include additional
information not recognised close to the cluster centre, these
discrepancies are likely explained by the difference in
field-size between the two studies. By mapping the mass distribution
towards a radius more than twice the size as that of \citet{dki02},
our results are better constrained. Further bias also arises from the
measurements of \citet{dki02} starting from an inner radius of
\mbox{$r_{\rm min} = 50\arcsec\; (\sim0.37\; h_{70}^{-1}{\rm Mpc})$},
where we consider the cluster galaxy contamination to be very high, in
addition to not including any correction for this contamination.

Our shear values are measured from a radial cut-off,
\mbox{$r_{\rm min}=150\arcsec$}, to avoid the large cluster galaxy
contamination present at small radii. Because $c$ is estimated from
the scale radius, $r_s=r_{200}/c$, it is desirable to include $r_s$ in
the measurements (\mbox{$r_{\rm min}<r_s$}) in order to obtain an
accurate estimate of the concentration parameter. If this is not the
case, $c$ is basically unconstrained.

This appears to be the case for Abell~1351, explaining why we
were not able to derive an upper limit for its concentration
parameter. Letting \mbox{$r_{\rm min}=150\arcsec$}, we ensure a
cluster galaxy contamination $<25$\% at this inner radius. However, as
the $c$ parameter appears unconstrained under this condition, we
reduce $r_{\rm min}$ attempting to obtain clearer results. The problem
now arising is the increasing contamination of cluster
galaxies. Looking at
Fig.~\ref{clusgalcont} we see that at \mbox{$r=120\arcsec$} the
cluster contamination is $\sim32$\%, and at \mbox{$r=100\arcsec$} it
equals $\sim40$\%. Though this contamination is accounted for during
the fitting process, the contamination correction is still vulnerable
to fluctuations in the projected galaxy density caused by foreground
and/or background structures. 

Table~\ref{cvary} presents the results from letting 
\mbox{$100\arcsec\leq r_{\rm min}\leq 150\arcsec$} for Abell~1351
(with the KS93 $\kappa$ map peak as cluster centre). It
is seen that whilst $c$ is decreasing with smaller $r_{\rm min}$,
$r_{200}$ and $M_{200}$ remain stable for different
$r_{\rm min}$. Also worth noticing is that for
\mbox{$r_{\rm min}\leq130\arcsec$}, $c$ becomes
constrained. However, as $r_{\rm min}>r_s$ for the different starting
radii, we cannot obtain further conclusions from these results. As
$r_s$ is even smaller for Abell~1995, we did not repeat this test for
the cluster. \citet{dsc05} experience similar problems when
attempting to determine the concentration parameter for Abell~222 and
Abell~223, concluding that obtaining a reliable $c$ from weak lensing
data only is difficult, if not impossible.

\begin{table*}
  \caption{Results from varying the inner radius from where the shear
    values of Abell~1351 are measured.}
  \label{cvary}
  \centering
  \begin{tabular}{ c c c c c c c }
    \hline \hline
    {$r_{\rm min}$}& {$r_s$}& {$c$}& {$r_{200}$}& {$M_{200}$}& No. of &{$\chi^2/{\rm Dof}$} \\
    {[$\arcsec$]}& {[$\arcsec$]}&& {[$h_{70}^{-1}$Mpc]}& {[$10^{14}h_{70}^{-1}M_\odot$]}& galaxies &\\ 
    \hline
    {$100$}& $76{+56\atop -35}$& {$4.9{+3.6\atop -2.2}$}& {$1.76{+0.057\atop-0.13}$}& {$8.5{+0.8\atop-1.9}$}& {$15\,630$} &$1.11$ \\
    {$110$}& $61{+48\atop -29}$& {$6.0{+4.7\atop -2.8}$}& {$1.71\pm0.10$}& {$7.9\pm1.4$}& {$15\,582$} &$1.45$ \\
    {$120$}& $54{+60\atop -26}$& {$6.8{+7.6\atop -3.3}$}& {$1.71{+0.10\atop-0.086}$}& {$7.9{+1.4\atop-1.2}$}& {$15\,529$} &$1.31$ \\
    {$130$}& $60{+83\atop -26}$& {$6.2{+8.5\atop -2.6}$}& {$1.76{+0.057\atop-0.11}$}& {$8.5{+0.8\atop-1.7}$}& {$15\,482$} &$1.33$ \\
    {$140$}& $37{+\infty\atop -20}$& {$10.0{+\infty\atop -5.5}$}& {$1.73{+0.11\atop-0.13}$}& {$8.1{+1.6\atop-1.8}$}& {$15\,428$} &$1.16$ \\
    {$150$}& $33{+\infty\atop -15}$& {$11.2{+\infty\atop -4.9}$}& {$1.74{+0.11\atop-0.10}$}& {$8.3{+1.6\atop-1.4}$}& {$15\,358$} &$0.96$ \\
    \hline
  \end{tabular}
\end{table*}

\subsubsection{Best fit concentration parameter}

\citet{bks01} present dark matter halo simulations, attempting to find
a ``best fit concentration parameter'' applicable to all types of
halos. They find that for halos of the same mass, the concentration,
\mbox{$c_{\rm vir}\equiv r_{\rm vir}/r_s$}, can be given by
\mbox{$c_{\rm vir}\propto(1+z_{\rm d})^{-1}$}. This is contrary to
earlier beliefs that $c_{\rm vir}$ does not vary much with
redshift. Numerically simulated massive clusters typically have
concentration parameters $\sim4-5$ \citep{bks01}. 
This is within the limiting values for both clusters, although looking
at Fig.~\ref{gtvsradius}, the outcome from varying $c$ seems to better
follow the shear values of Abell~1351.

There exists several examples of large concentration parameters in the
literature. \citet{khe03} find \mbox{$c=22{+9\atop -5}$} for the
central mass concentration of the cluster
\mbox{${\rm Cl\;}0024+1654$}. \citet{gav05} conclude on
\mbox{$c=11.73\pm0.55$} for \mbox{${\rm MS}2137-23$}, while
\citet{btu05} find \mbox{$c=13.7{+1.4 \atop -1.1}$} for
Abell~1689. \citet{lrj06} present a thorough discussion of the
different concentration parameters derived for Abell~1689 in the
literature, and conclude that a distribution of best fit $c$
parameters is needed for observed lensing clusters in order to provide
a sample large enough to make an adequate comparison with
simulations. A recent study of observed concentration values for
clusters by \citet{cn07} show that the best fit lensing-derived $c$
parameters are systematically higher than concentrations derived via
X-ray measurements, a difference which can be at least partly
explained by effects of triaxiality of cluster halos
\citep{ck07,gav05,otu05,cdk04} or the sub-structure within the clusters
\citep{kc07}, although the latter effect may also produce a negative
bias of $c$ values.
In addition, baryonic physics can increase the concentration parameter 
mildly by up to $10\%$ as compared to dissipationless dark matter in pure
dark matter simulations \citep[see e.g.][]{ljm06}.

\subsection{Centre position}\label{centrepos}

In addition to the three centre positions tested in Sect.~\ref{model},
we computed $\kappa$ maps with the method of \citet[][KS93]{ks93} and
utilised the peak of this surface mass distribution as a fourth cluster
centre. The KS93 method assumes that $\gamma=g$, which is not a good 
approximation near the centres of massive systems. Therefore, in 
comparison with the other methods, it provides us with a reference 
point as for how large a variation one can reasonably expect for 
the various centroiding methods.

All centre positions obtained with the four methods lie within $1\arcmin$ 
and hence represent the errors expected when using the peak of a 
$\kappa$ map as cluster centre. As can be seen from Table~\ref{results}, 
varying the centre position only slightly (maximum distance between cluster
centres is $<55\arcsec$) can lead to different mass estimates. Although 
within error bars, the results from fitting NFW using a fixed $c$ varies
most. The NFW fitting of two parameters is more stable with a smaller 
spread in $M_{200}$. This is also reflected in $\chi^2/{\rm Dof}$, as a 
value closer to $1$ is a better fit.

Worth noticing is the generally smaller differences between the
results of Abell~1995 as compared to those of Abell~1351. The concentration
parameter seems also better constrained for Abell~1995, where we could not
obtain an upper limit for $c$ only in the case where the BCG was used as 
the centre reference. For Abell~1351 on the other hand an upper limit
for $c$ could not be obtained for any of the cluster centres chosen. 
This is consistent with the fact that Abell~1351 is not in dynamical 
equilibrium, lacking a well-defined cluster centre. The results obtained 
from fitting spherically symmetric models will hence depend on the centre 
chosen.

\subsubsection{Strong lensing features}\label{stronglensfeatures}
This picture of the centre of Abell~1351 is confirmed also by strong 
lensing effects, which is in general susceptive to sub-structures in 
clusters. For both clusters recent archival WFPC2 HST data exists, taken 
for an ongoing snapshot survey of X-ray luminous cluster (HST PID 11103, 
PI: H. Ebeling). The images are taken through the F606W filter totalling 
1200s exposure time each.

Considering both the colours and morphologies of galaxies in our
\mbox{$V-I$} data
and the morphologies in the HST images, there are at least half a dozen 
plausible arcs and arclets visible in each of the two clusters. The lensing 
pattern for Abell~1351 appears to be very complex and does not indicate a 
single, well-defined centre. This is supported by the presence of several 
elliptical galaxies which are of similar brightness as the BCG. 
On the contrary, for Abell~1995 several arc(let)s are well aligned around 
the BCG, apart from three which are obviously associated with individual 
cluster galaxies. Thus adopting the BCG for Abell~1995 as the centre is 
justified. Strong lensing will therefore not offer more constraints
on the determination of the centre of mass than we already have.

\subsection{The mass reconstructions}

In Sect.~\ref{mass} we presented the weak lensing reconstruction of
the clusters' surface mass density, detecting the clusters on the
$5\sigma$ level. In case of Abell~1351 a significant neighbouring peak
A was detected, which coincides with the galaxy distribution.

We note significant differences comparing our mass maps to those of
\citet{dki02}, who used the KS93 inversion method \citep{ks93}. Abell~1995
appears rather circular in both reconstructions, with attached filamentary
structures of low significance seen in the KS93 map. In their re-analysis of
the Abell~901 supercluster field \citet{hgp08} argue that such filamentary
structures could be enhanced by the KS93 method itself. This algorithm 
assumes $g=\gamma$ near the critical cluster cores, which hence overestimates
$\kappa$, and the smoothing implemented could then lead to the apparent
merging of neighbouring peaks.

Our map of Abell~1351 appears roughly circular at the position of the main
cluster with a significant extension towards the north-east, whereas it shows
an extension to the south-west in the work of \citet{dki02}. Both
reconstructions have peak B in common. Since it is not associated with any
over-density of galaxies but appears in both reconstructions based on very
different data sets, the most likely explanation is a chance alignment of
galaxies triggering this detection.

\section{Conclusions}\label{conclusion}

Utilising observations from CFH12K we find the masses of Abell~1351
(Abell~1995) to be
$M_{200}\equiv M(r_{200})\sim8-9\times10^{14}\;h_{70}^{-1}M_\odot$
($M_{200}\sim5-6\times10^{14}\;h_{70}^{-1}M_\odot$). These
results are derived from fitting our data to the NFW profile, altering
both $c$ and $r_{200}$, to find the best theoretical fit to our shear
measurements (Fig.~\ref{gtvsradius}). Our $\kappa$
reconstructions yield mass estimates of
$M_{\rm 2D}(<r_{200})=11.7\pm3.1\times10^{14}h_{70}^{-1}M_\odot$
($M_{\rm 2D}(<r_{200})=10.5\pm2.7\times10^{14}h_{70}^{-1}M_\odot$) for
Abell~1351 (Abell~1995).

Our results illustrate that using solely weak lensing measurements,
with no photometric or spectroscopic information to separate cluster
members from background galaxies, the concentration parameter found
for a galaxy cluster is poorly constrained. Future spectroscopic
redshift measurements and strong lens modelling of the arcs seen
towards the cores of Abell~1351 and Abell~1995 might help improve the
constraints on their concentration parameters. However, the change in
$c$ value resulting from small variations of the centre position of
Abell~1351 indicates that the mass distribution in the core of this
dynamically unrelaxed cluster is too complex to be meaningfully fit by
the NFW mass density profile.
$M_{200}$, on the other hand, varies only slightly for both
clusters when either $r_{\rm min}$ is decreased or the cluster
centre varied. Although the KSB+ PSF correction leads to an
underestimation of the cluster masses (with a maximum of $20\%$ in
our case), increasing the background galaxy ellipticities
correspondingly still yields masses within error bars of our current
estimates. We therefore conclude that the mass estimates presented
are robust.

\begin{acknowledgements}
The authors thank the anonymous referee for very helpful critics, and
Thomas Erben and Peter Schneider for useful and constructive
comments. We also thank Aleksi Halkola for helpful discussion during
the NFW analysis. K.H. gratefully acknowledges support from a doctoral
fellowship awarded by the Research council of Norway, project number
177254/V30.

\end{acknowledgements}

\bibliography{ref}
\end{document}